\documentclass[aps,gbroupedaddress,amsmath,amssymb]{revtex4-2}

\usepackage{anyfontsize} 

\usepackage {enumerate}
\usepackage {float}
\usepackage{epstopdf}
\usepackage{xcolor}
\usepackage{CJKutf8} 
\usepackage{ulem}
\usepackage[breaklinks=true]{hyperref}
\usepackage{setspace}

\usepackage{amssymb}
\usepackage{stmaryrd}
\usepackage{amsmath}
\usepackage{amsfonts}
\usepackage{mathrsfs}
\usepackage[utf8]{inputenc}
\usepackage{amsmath,amssymb,amsfonts}
\usepackage{graphicx}
\usepackage{subfigure}
\usepackage{color} 
\usepackage{fancyhdr}
\usepackage{hyperref} 
\begin{document}
\begin{CJK}{UTF8}{gbsn}
	
	\title{Analytic solutions for the motion of spinning particles near brane-world black hole}
	
	\author{Yunlong Liu}
	\affiliation{Department of Physics, South China University of Technology, Guangzhou 510641, China}
	\author{ Xiangdong Zhang\footnote{Corresponding author. scxdzhang@scut.edu.cn}}
	\affiliation{Department of Physics, South China University of Technology, Guangzhou 510641, China}


	\begin{abstract}
		The general motion of spinning test particles to the leading order approximation of spin in the brane-world spacetime is investigated. Analytical integrations for the equations of motion and linear shifts in orbital frequency are obtained. As a result, we found that both the nodal precession and the periastron precession become larger when the tidal charge $b$ of brane-world spacetime becomes smaller. For the periastron precession, the effect is further amplified as the spin increases. Our work can potentially be applied to the study of gravitational waveforms of Extreme Mass Ratio Inspirals with spin in brane-world spacetime.
	\end{abstract}
	
	\maketitle

	\section{Introduction}\label{Intro}
	
	It has been nearly 10 years since LIGO first directly detected gravitational waves(GWs) \cite{Tests_ligo_2016}, ground-based GW detection technology has been continuously updated and iterated, and the number of observed events has steadily increased \cite{GWTC-2_LIGO_2021, GWTC-3_LIGO_2023, GWTC-2_LIGO_2024}. More notably, space-based GW detectors such as LISA \cite{Laser_Amaro-Seoane_2017}, Taiji \cite{Taiji_Hu_2017, Alternative_Wang_2021}, and TianQin \cite{Science_Fan_2020} are also being developed intensively. These detectors can extend the sensitivity range to lower frequencies. It  means that they could observe a special class of binary systems known as Extreme Mass Ratio Inspirals (EMRIs) \cite{Science_Babak_2017} that describe a system where a small secondary object orbits a supermassive  primary one. Analyzing the GW signals from EMRIs allows us to explore the nature of compact objects at the center of galaxies as well as to test the General relativity(GR) \cite{New_Arun_2022}.
	
	Modeling the gravitational waveforms of EMRIs typically involves expansion in the mass ratio. In the leading order, adiabatic methods for EMRIs encodes the spiral  inward of the secondary object that caused by GW fluxes along with the geodesic \cite{Gravitational_Hughes_2005,Adiabatic_Hughes_2021},  while the first post-adiabatic order includes conservative self-force effects and the spin of the test body \cite{Gravitational_Wardell_2023}.  These post-adiabatic effects, particularly the spin of the test particle, have a crucial impact on the data analysis and modeling of gravitational waveforms \cite{Importance_Huerta_2012, Assessing_Piovano_2021}.
	To further explore the dynamics of EMRI systems with spins, there are primarily two approaches: the post-Newtonian(PN) approximation schemes \cite{Precessional_Gerosa_2015, Multi-timescale_Gerosa_2015, Effective_Kesden_2015, Endpoint_Mould_2020, Integrability_Tanay_2021, Spin-eccentricity_Fumagalli_2023, Action-angle_Tanay_2023} and effective-one-body(EOB) schemes \cite{Coalescence_Damour_2001, Effective-one-body_Balmelli_2013, Effective-one-body_Balmelli_2015}. At the same time, many papers have studied the effects of spin on physics such as the innermost stable orbits(ISO) \cite{Conservative_Favata_2011}, collisional Penrose processes \cite{Collisional_Mukherjee_2018}, black hole accretors \cite{Kerr_GuoMY_2016}, gravitational waveforms \cite{Prospects_Rahman_2023}. However, since test body's spin couples to the curvature of that spacetime, we need to consider the spin-curvature force and must be modeled carefully in order to accurately characterize the motion of bodies orbiting black holes. In order to have a better understanding on these important issues. Recently, an analytic solution for the motion of spinning particles to leading order in spin in the static and spherically symmetric spacetimes has been provided in \cite{Analytic_Witzany_2024} along  \cite{Periastron_Hinderer_2013,Precisely_Drummond_2022, Precisely_Drummond_2022a}. As a result, they obtain the bound orbits in Schwarzschild space-time by expressing the solution in the form of Jacobi elliptic functions.

	On the other hand, to unify gravity with other fundamental forces, an elegant scheme is the extra dimensions theory, such as the Kaluza-Klein (KK) model\cite{Quantum_Klein_1926,Unification_Kaluza_2018} and the later the string/M-theory, both hypothesize the existence of extra dimensions at the Planck scale, which are currently difficult to detect with experiments.
	Differently, as an effective four-dimensional version of high-dimensional string theory, the brane-world model\cite{Einstein_Shiromizu_2000, Brane-World_Maartens_2010} posits that our physical universe is embedded in a higher-dimensional spacetime (bulk), with standard model matter and fields trapped on the brane. Interestingly, a black hole solution can be obtained in the brane-world model \cite{Black_Dadhich_2000}, characterized by the tidal charge $b$.
	This in turn provides a framework to study the physical effects caused by extra dimensions \cite{Bounds_Nucamendi_2020, Periodic_Deng_2020, Collisional_Du_2022, Testing_Stojiljkovic_2023}.
	
	Given the above motivation and considering the importance of analytic solutions. In this paper, we aimed to generalize the analytic solutions obtained in \cite{Analytic_Witzany_2024} to brane-world spacetime and discusses the physical consequence of the tilde charge parameter $b$.
	
	The structure of this article is as follows:
	In Sec. \ref{SandW}, the background spacetime is introduced  and the corresponding equations of motion are provided. Sec. \ref{PT} introduces parallel transport of spin and provides the evolution of spin components. In Sec. \ref{PT}, the Kepler-parameter expressions of  energy, angular momentum, and the roots of radial velocity are given. Sec. \ref{Analytic} and \ref{FShift} respectively presented analytic results of motion and linear shifts in orbital frequencies. Finally, in sec. \ref{conclusion}, some conclusions and corresponding discussions are provided. Through out the paper, the geometrized units with $G=c=1$ and the metric signature ($-+++$) are used.
	
	\section{Equations of motion}\label{SandW}
	
	\subsection{Brane-world background}
	The field equations on the  bulk(5 dimensions) and the reduced equations of motions on brane(4 dimensions) is given respectively as \cite{Einstein_Shiromizu_2000}
	\begin{eqnarray}
		G^{(5)}_{A B} &=& - \tilde{\Lambda} g^{(5)}_{A B} + \tilde{\kappa}^{2} T^{(5)}_{A B},\\
		G_{ \mu \nu } &=& - \Lambda g_{\mu\nu} + \kappa^{2} T_{\mu\nu} +   \tilde{\kappa}^{4} \mathcal{S}_{\mu\nu} - \mathcal{T}_{\mu\nu},
	\end{eqnarray}
	where $\tilde{\Lambda}$ is  the 5-dimensional negative cosmological constant, $T^{(5)}_{A B}$ represents 5-dimensional energy-momentum, and $\tilde{\kappa}^2 =  8 \pi/ \tilde{M}_p^3$ with $\tilde{M}_p$ being 5-dimensional Planck mass in the bulk. Meanwhile, $\kappa^2 = 8\pi / M_p^2=\lambda \tilde{\kappa}^4/6$ and $\Lambda = \tilde{\kappa}^{2} ( \tilde{\Lambda} + \tilde{\kappa}^2 \lambda^2)/2$   where $\Lambda$, $M_p$ and $\lambda$ represent the cosmological constants, effective Planck mass and brane tension  in the brane, respectively. Here, we just consider the case where $ \tilde{\Lambda} = - \tilde{\kappa}^2 \lambda^2$. $T_{\mu\nu}$ and  $\mathcal{S}_{\mu\nu}$ are the energy momentum and its quadratic parts, separately. Moreover, $\mathcal{T}_{\mu\nu}$ is the projected Weyl tensor on the brane \cite{Einstein_Shiromizu_2000}. 
	
	We usually assume that there is no exchange of energy-momentum between the bulk and  brane in order to obtain conservation equations \cite{Brane-World_Maartens_2010} as
	\begin{eqnarray}
		\nabla^\nu T_{\mu\nu} = -2 T^{(5)}_{AB}  n^A g^B_\mu = 0, 
	\end{eqnarray}
	where $\nabla_\nu$ is the covariant derivative on the brane. 
	This means that, we can naturally use the ``dynamical equation'' of spin particles  with only changing the background metric in the brane-world. 
	
	On the other hand, we have
	\begin{eqnarray}
		\nabla^\nu   \mathcal{T}_{\mu\nu} = \tilde{\kappa}^{4}  \nabla^\nu \mathcal{S}_{\mu\nu}.
	\end{eqnarray}
	For test bodies, in the low-energy approximation \cite{Einstein_Shiromizu_2000,Brane-World_Maartens_2010}, we assume that the quadratic energy-momentum tensor $\mathcal{S}_{\mu\nu}$ has a negligible impact on the background spacetime. It  leads to $\nabla^\nu   \mathcal{T}_{\mu\nu} \approx 0$. Hence, We can apply the conclusion of \cite{Black_Dadhich_2000} and obtain the metric of the brane-world \cite{Black_Dadhich_2000}: 
	\begin{eqnarray}
		ds^2 = - f[r] dt^2 + \frac{1}{f[r]}dr^2  + r^2\left( d\theta ^2 + \sin^2\theta  d\phi ^2 \right), \label{RNMertic}
	\end{eqnarray}
	where,$f[r] =1- 2 M/ r +  M^2 b/r^2 $. Here, $M$ is the mass of the black hole, and $b$ is the tidal charge parameter arising from the bulk Weyl field.
	When $f[r] = 0$ , corresponding to two horizons of the black hole:
	\begin{equation}
		r_n = M \left( 1 - \sqrt{ 1 -b} \right),\quad
		r_p = M \left( 1 + \sqrt{ 1 -b} \right),
	\end{equation}
	where, the range of $b$ is constrained as $b \leq 1$.
	
	\subsection{Leading order in small spin}\label{MPD}
	
	In curved spacetime, any distribution of mass or other properties will cause the particle to move away from the geodesic. For a spinning test body, in brane world spacetime, the motion can be described by the famous Mathission–Papapetrou–Dixon
	(MPD) equations as \cite{SpinningI_Papapetrou_1951, Dynamics1_Dixon_1970, Dynamics2_Dixon_1970}:
	
	\begin{eqnarray}
		\frac{DP^\mu}{d \tau}& = &-\frac{1}{2} R^{\mu}{}_{\nu \alpha \beta} \dot{x}^{\nu}S^{\alpha \beta},\\
		\frac{DS^{\mu \nu}}{d \tau}&=&P^\mu \dot{x}^\nu-P^\nu \dot{x}^\mu,
	\end{eqnarray}
	where $\dot{x}^\mu=dx^\mu/dt$, $P^\mu$ and $S^{\mu\nu}$ respectively represent tangent vector to
	the worldline of test particle, 4-momentum and the spin angular momentum tensor.

	The MPD equations is not sufficient for the evolution of all degrees of freedom. Hence, spin supplementary condition (SSC) should be required. We commonly adopt the form proposed by Tulczyjew-Dixon (TD) \cite{Motion_W_1959,Dynamics1_Dixon_1970}, given by $S^{\mu \nu}P_{\nu}=0$.  With this condition, the four-velocity $\dot{x}^\mu$ and the four-momentum $P^\mu$ are not parallel. However, if we consider only the linear order of spin, we can obtain:
	\begin{eqnarray}
		-P_{\mu} \dot{x}^\mu &=& m + \mathcal{O}[S^2],\\
		P^{\mu} &=& m \dot{x}^\mu + \mathcal{O}[S^2].
	\end{eqnarray}
	Here, $m$ represents the mass of the spinning test body.
	Similarly, the spin tensor is antisymmetric,  and approximately expressed as:
	\begin{eqnarray}
		S^{\alpha \beta} = m \epsilon^{\alpha \beta \mu \nu}\dot{x}_\mu  s_\nu  + \mathcal{O}[S^2],
	\end{eqnarray}
	where $s^\nu$ is the specific spin vector. Meanwhile, the TD condition transforms into $s^{\nu}\dot{x}_{\nu}=0$. This implies that  $s^{\nu}$ is a spatial vector.
	
	Finally, the MPD equations of motion can be reformulated as:
	\begin{eqnarray}
		\frac{D^2 x^\mu}{d \tau^2}&=&-\frac{1}{2}  R^{\mu}{}_{\nu \alpha \beta} \epsilon^{\alpha \beta}{}_{\lambda \sigma} \dot{x}^{\nu} \dot{x}^\lambda  s^\sigma , \label{DDx}\\ 
		\frac{D s^{\mu}}{d \tau}&=& 0. \label{Ds}
	\end{eqnarray}
	It can be seen that at the linear order, the spin vector undergoes parallel transport along the world line.
	
	\subsection{Equations of motion}\label{motionEqs}
	In spherically symmetric spacetime, there exist Killing vector fields
	\begin{eqnarray}
		\xi_{(t)\mu} &=& \{1,0,0,0\},\\
		\xi_{(x)\mu} &=& \{0,0, -\sin \phi,- \cos \phi \cot \theta \},\\
		\xi_{(y)\mu} &=& \{0,0,\cos \phi, -\sin \phi \cot \theta \},\\
		\xi_{(z)\mu} &=& \{0,0,0,1\}.
	\end{eqnarray}
	These Killing vector fields correspond to time translations and spatial rotations, respectively.
	
	Correspondingly, the conserved quantity for motion with the given Killing vector field in the brane-world spacetime reads
	\begin{eqnarray}
		C_{(\xi)} = P^\mu \xi_\mu -\xi_{\rho;\sigma} S^{\rho \sigma}/2 .
	\end{eqnarray}
	Using the  Killing vector fields, the energy and angular momentum of a test particle can be described as:
	\begin{eqnarray} \label{EJConstant}
		\mathcal{E} &=& f\dot{t} +  \frac{1}{2} r^2 \sin \theta \left(s^\phi \dot{\theta} - s^\theta \dot{\phi}\right) f'[r],\label{EKilling}\\
		\mathcal{J}_x &=&  r^{2} \left( -\dot{\phi} \cos\theta \cos\phi \sin\theta - \dot{\theta} \sin\phi  \right) + \cos \phi \sin \theta \left(\dot{r} s^{t} - \dot{t} s^{r} \right)   \notag \\
		&&+ r f[r] \left(\cos \theta \cos \phi \left(\dot{\theta}  s^{t}−\dot{t} s^{\theta}\right) + \sin\theta \sin\phi  \left(−\dot{\phi} s^{t} + \dot{t} s^{\phi}\right)\right), \label{mathcalJx}\\
		\mathcal{J}_y &=& r^{2} \left( \dot{\theta} \cos\phi - \dot {\phi}  \cos \theta \sin\theta \sin\phi \right) + \sin\theta \sin\phi \left( \dot{r} s^{t} - \dot{t} s^{r} \right)   \notag \\ 
		&& + r f[r] \left( \cos\theta \sin\phi \left( \dot{\theta} s^{t} - \dot{t} s^{\theta} \right) + \cos\phi \sin\theta \left( \dot {\phi} s^{t} - \dot{t} s^{\phi}\right) \right), \label{mathcalJy}\\
		\mathcal{J}_z &=& r^{2} \sin^{2}\theta + \cos\theta \left( \dot{r} s^{t} - \dot{t} s^{ 1} \right) +  r f[r]\sin\theta \left(-\dot{\theta} s^{t} + \dot{t} s^{\theta} \right). \label{mathcalJz}
	\end{eqnarray}
	Here, $\mathcal{E}$, $\mathcal{J}_x$, $\mathcal{J}_y$, and $\mathcal{J}_z$ are all normalized by the test particle's mass $m$. 
	We can always reselect the aligned coordinate system such that the angular momenta $\mathcal{J}_x = 0$, $\mathcal{J}_y = 0$, and $\mathcal{J}_z = \mathcal{J}$. Then, we assume the $\theta$ satisfies $\theta = \pi/2 + \delta\theta$. Substituting into Eqs. \eqref{mathcalJx}-\eqref{mathcalJy}  and expanding in terms of $\delta\theta$, we obtain 
	\begin{eqnarray} \label{DJxJy}
		\mathcal{J}_{x} &=& -\sin\phi \left( \dot{\delta\theta} r^{2} + r f[r] \left(\dot{\phi}  s^{t} -\dot{t} s^{\phi} \right) \right) + \cos\phi \left( \dot {r} s^{t} - \dot{t} s^{r} + \dot{\phi} r^{2} \delta \theta \right) ,\\
		\mathcal{J}_{y} &=& \cos\phi \left( \dot{\delta\theta} r^{2} + r f[r] \left( \dot{\phi} s^{t} - \dot{t} s^{\phi} \right) \right) + \sin\phi \left( \dot{r}s^{t}- \dot{t}s^{r}+ \dot{\phi}r^{2}\delta \theta \right).
	\end{eqnarray}
	To make $\mathcal{J}_x = 0$ and $\mathcal{J}_y = 0$, $\delta\theta$ and its time derivative need to satisfy
	\begin{eqnarray} \label{Dtheta}
		\dot{\delta\theta}&=&-\frac{1}{r}f[r] \left(\dot{\phi}  s^{t} -\dot{t} s^{\phi} \right) + \mathcal{O}[s^2],\\
		\delta \theta &=& -\frac{\dot{r} s^{t} - \dot{t} s^{r} }{ \dot{\phi} r^{2}} + \mathcal{O}[s^2]. \label{deltatheta}
	\end{eqnarray}
	Similarly, substituting $\theta = \pi/2 + \delta\theta$ into $\mathcal{E}$ and $\mathcal{J}_z$ and expanding in terms of $\delta\theta$, and make use of Eq.\eqref{Dtheta}, we obtain
	\begin{eqnarray} \label{DJzE}
		\mathcal{E} &=&f[r] \dot{t} + \frac{1}{2}r^{2} f'[r]  \dot{\phi} s^{\theta} + \mathcal{O}[s^2] ,\\
		\mathcal{J}&=& \dot {\phi} r^{2} +  r f[r]\dot{t} s^{\theta} + \mathcal{O}[s^2]  .
	\end{eqnarray} By employing the normalization condition $P_{\mu}P^{\mu}=-m^2$, we further obtain
	\begin{eqnarray}
		\dot{t} &=&  \frac{\mathcal{E}}{f[r]} + \frac{\mathcal{J}f'[r] s_{\parallel}}{2 r f[r] }, \label{dtdt}\\
		\dot {\phi} &=& \frac{\mathcal{J}}{ r^{2}} + \frac{\mathcal{E} s_{\parallel}}{r^2} ,\\
		\dot {r}^{2} &=& \mathcal{E}^2  - \frac{(\mathcal{J}^{2} + r^2) f[r]}{r^{2}} - \frac{\mathcal{J} \mathcal{E} ( 2f[r] - rf^{\prime}[r] ) s_{\parallel}}{r^2}=\mathcal{R}[r]. \label{drdt}
	\end{eqnarray}
	Here, the conclusion in later Eq.\eqref{stheta} is used. Note that Eq.\eqref{stheta} uses only the part of Eq.\eqref{dtdt}-\eqref{drdt} without spin parameters, so we can use the conclusion of Eq.\eqref{stheta} with confidence. At the same time, the orbital precision is $\mathcal{O}[s^2]$. Note that Equation \eqref{drdt} and Eq.(17) in the Ref.\cite{Analytic_Witzany_2024} differ by a negative sign in the $s_\parallel$ term.
	
	In subsequent calculations, we will use parameters normalized by the black hole mass $M$, such as $\bar{r}=r/M$, $\bar{\mathcal{J}}=\mathcal{J}/M$, and $\bar{s}_\parallel=s_\parallel/M$. Moreover, in the following, we will omit the overline notation $\bar{\ }$ for simplicity. For example, $f[r]$ will become $f[r]=1-2/r+b/r^2$, and $r_p$ will be abbreviated as $r_p=1+\sqrt{1-b}$.
	
	\section{Parallel transport}\label{PT}
	From Eq.\eqref{Ds}, it can be seen that the spin is parallel transported along the orbit. Additionally, in the equations of motion, it was found that the influence of  spin on the evolution of $\{t, r, \theta, \phi\}$ is decoupled. For example, only the component of the $s_\theta$ affects the evolution of $\{t, r, \phi\}$. Based on this, the Ref. \cite{Analytic_Meent_2020} further expanded the usage of the Marck tetrad \cite{Solution_Marck_1983} and obtained a closed-form tetrad to depicted the parallel transport. Along this line, the Ref.\cite{Precisely_Drummond_2022a, Precisely_Drummond_2022} calculated the orbits of test particles with small spin and analyzed how the spin of small objects affects important observables, such as orbital frequencies.
	
	Under the guidance of  4-velocity and orbital angular
	momentum, as the zeroth and second leg of the tetrad \cite{Precisely_Drummond_2022a}
	\begin{eqnarray} 
		e^\mu_{(0)} &=& \left\{ \frac{\mathcal{E}}{f[r]} , \dot{r}, 0, \mathcal{J}/r^2 \right\} +\mathcal{O}[s], \\
		e^\mu_{(2)} &=& \left\{0,0,-1/r,0\right\} +\mathcal{O}[s].
	\end{eqnarray}The direct calculation shows
	\begin{eqnarray} 
		\frac{d e^\mu_{(0)}}{D\tau} &=& \{0,0,0,0\} + \mathcal{O}[s] ,
		\\
		\frac{d e^\mu_{(2)}}{D\tau} &=& \{0,0,0,0\} + \mathcal{O}[s] .
	\end{eqnarray}
	This confirms that zeroth and second leg of the tetrad are parallel transported along the worldline. On the other hand, we can choose  the remaining  two  legs that are orthogonal to the zeroth and second legs. According to  Ref.\cite{Solution_Marck_1983, Analytic_Meent_2020}, the other two legs can be defined as:
	\begin{eqnarray} 
		\hat{e}^\mu_{(1)} &=& \left\{ \frac{\dot{r} r}{f[r] \sqrt{\mathcal{J}^{2} + r^{2}}}, \frac{\mathcal{E}r}{ \sqrt{\mathcal{J}^{2} + r^{2}}}, 0 , 0 \right\} +\mathcal{O}[s],
		\\
		\hat{e}^\mu_{(3)} &=& \left\{ \frac{\mathcal{J} \mathcal{E} }{ f[r] \sqrt{\mathcal{J}^{2} + r^{2}}}, \frac{\mathcal{J} \dot{r}}{ \sqrt{\mathcal{J}^{2} + r^{2} }} , 0 , \frac{ \sqrt{ \mathcal{J}^{2} + r^{2}}}{ r^{2}}\right\} +\mathcal{O}[s], \label{hatemu3F}
	\end{eqnarray}
	Note that Eq.\eqref{hatemu3F} and Eq.(23) in Ref. \cite{Analytic_Witzany_2024} are not consistent, but it matches their Supplemental Materials. 
	Note that the legs $\hat{e}^\mu_{(1)}$ and $\hat{e}^\mu_{(3)}$ are not parallel transported with a spin precession. Therefore, these two legs should be re-parameterized using the parameter $\psi$ as
	\begin{eqnarray} 
		e^\mu_{(1)} &=&  \cos\psi \hat{e}^\mu_{(1)}  -\sin\psi \hat{e}^\mu_{(3)} ,
		\\
		e^\mu_{(3)} &=&  \sin\psi \hat{e}^\mu_{(1)}  +\cos\psi \hat{e}^\mu_{(3)}.
	\end{eqnarray}
	Through the parallel transport of $e^\mu_{(1)}$ and $e^\mu_{(3)}$ along the geodesic, we obtain
	\begin{eqnarray} 
		\frac{d\psi[r] }{d\tau}= \frac{\mathcal{J} \mathcal{E}}{\mathcal{J}^2+r^2}. \label{psitau}
	\end{eqnarray} In the Marck tetrad, we have
	\begin{eqnarray} 
		\frac{d (s_\mu e^\mu_{(\rho)})}{D\tau} =s_\mu\frac{d  e^\mu_{(\rho)}}{D\tau}+ e^\mu_{(\rho)} \frac{d s_\mu}{D\tau} = 0 + \mathcal{O}[s^2].
	\end{eqnarray}
	Therefore, the spin components in the Marck tetrad $\{s^t_M, s^r_M, s^\theta_M, s^\phi_M\}$ are constants at $\mathcal{O}[s^2]$ order. Additionally, from the TD condition mentioned earlier, we know that $s^t_M = 0$. Through direct calculations, we obtain the following relations:
	\begin{eqnarray} 
		s^{t} &=& \frac{\mathcal{J} \mathcal{E}\left(-\sin[\psi] s^r_M + \cos[\psi] s^\phi_M \right) + \dot{r}r  \left(\cos[\psi]s^r_M + \sin[\psi] s^\phi_M\right)}{ f[r]\sqrt{J^{2}+ r^{2}} },
		\\
		s^{r}&=& \frac{ \mathcal{J} \dot{r}  \left( -\sin[\psi]s^r_M+ \cos[\psi]s^\phi_M\right) + \mathcal{E}r \left( \cos[\psi]s^r_M+ \sin[\psi]s^\phi_M\right)}{ \sqrt{\mathcal{J }^{2}+ r^{2}} },\\
		s^{\theta}&=& - \frac{s^\theta_M}{r }, \label{stheta}\\
		s^{\phi}&=& \frac{\sqrt{\mathcal{J }^{2}+ r^{2}} \left(-\sin[\psi]s^r_M+ \cos[\psi]s^\phi_M\right)}{r^{2}}.
	\end{eqnarray} Note that we can always reselect the parameters $\hat{\psi}=\psi + \arg[s_M^{r} +  i s_M^{\phi}]$, $s_\perp=\sqrt{ (s_M^{r})^2 + (s_M^{\phi})^2}$ and $s_\parallel=s^\theta_M $ such that
	\begin{eqnarray} 
		\sin[\hat{\psi}] s_\perp  &=& \left(\cos[\psi]s_M^{r} + \sin[\psi]s_M^{\phi}\right),\\
		- \cos[\hat{\psi}] s_\perp &=& \left(\sin[\psi] s_M^{r} - \cos[\psi] s_M^{\phi} \right).
	\end{eqnarray}
	
	Finally, the evolution of each component of the spin vector can be simplified to
	\begin{eqnarray} 
		s^{t} &=& \frac{(\mathcal{E} \mathcal{J}\cos[\hat{\psi}] + \dot{r} r \sin[\hat{\psi}] ) s_\perp}{f[r] \sqrt{\mathcal{J }^{2}+ r^{2}} },\label{st}
		\\
		s^{r} &=& \frac{( \mathcal{J} \dot{r} \cos[\hat{\psi}]  + \mathcal{E}r \sin[\hat{\psi}] ) s_\perp}{\sqrt{\mathcal{J }^{2}+ r^{2}} },\label{sr}\\
		s^{\theta}&=& - \frac{s_\parallel}{r },\\
		s^{\phi} &=& \frac{\cos[\hat{\psi}] \sqrt{\mathcal{J }^{2}+ r^{2}} s_\perp }{ r^{2}}
	\end{eqnarray}
	
	Combining the above results with Eq.\eqref{deltatheta}, we can obtain 
	\begin{eqnarray} 
		\delta \theta = \frac{s_\perp \sin[\hat{ \psi}] \sqrt{\mathcal{J}^2+r^2}}{\mathcal{J} r} +\mathcal{O}[s^2 ].
	\end{eqnarray}
	
	\section{Kepler parametrization}\label{KP}
	In subsequent calculations, we only consider the bound orbits between the two radial turning points $[r_2, r_1]$. Therefore, we can always use the Keplerian parameters $\{e, p\}$ to re-describe the approximate expressions for the energy $\mathcal{E}$ and angular momentum $\mathcal{J}$ \cite{gravity_Darwin_1959}. The transformation relations are as follows:
	\begin{eqnarray}
		r_1=\frac{p}{1-e},\quad r_2=\frac{p}{1+e}.  \label{r12}
	\end{eqnarray}
	Here, $r_1$ and $r_2$ are the two real roots of $\mathcal{R}[r]$ defined in Eq.\eqref{drdt}, corresponding to the apoapsis and periapsis of the radial motion, respectively. 
	
	\subsection{Energy  and angular momentum}
	Reorganizing $\mathcal{R}[r]$, we obtain
	\begin{eqnarray}
		\mathcal{E}^{2}- f[r] - F _{\mathrm { R}2}[r] \mathcal{J}^{2} + \mathcal{J} \mathcal{E} F_{\mathrm{R}3}[r] s_\parallel= 0, \label{Rr}
	\end{eqnarray}
	where, 
	\begin{eqnarray}
		F_{\mathrm{R}2}[r]&=& \frac{f[r]}{ r^{2}}, \\
		F_{\mathrm{R}3}[r]&=& \frac{- 2 f[r]+ r f^{\prime}[r]}{ r^{2}}.
	\end{eqnarray}
	Because $s_\parallel$ is small, we first neglect the small part $F_{\mathrm{R}3}[r]$ and solve  the equations $\mathcal{R}[r_{1}] = 0$ and $\mathcal{R}[r_{2}] = 0$ to obtain the leading-order solutions $\{\mathcal{E}_b^2, \mathcal{J}_b^2\}$. Then, using perturbation schemes, we derive the next-order correction terms involving $s_\parallel$. This gives us the relationship between $\{\mathcal{E}^2, \mathcal{J}^2\}$ and $\{e, p\}$ as follows:
	\begin{widetext}
		\begin{eqnarray}
			\mathcal{E}^{2} &=&\frac{p^2 Y-2 b p (p (Z-2)+2 Z)+ b^2 Z^2 }{p^2 V} +s_\parallel Z^2 \left(p^2+b (Z-2 p)\right)  \frac{ \sqrt{(p-b)W}}{p^3 V^2},  \label{E2toep}\\
			\mathcal{J}^{2} &=& \frac{p^2 (p-b)}{V}  - s_\parallel (p (2 p+3 Z - 12)+8 b (2-Z)) \frac{\sqrt{(p-b) W }}{V^2},  \label{J2toep}
		\end{eqnarray}
	\end{widetext}
	where, 
	\begin{eqnarray}
		X &=& p-3-e^2,\\
		Y&=& (p-2)^2-4 e^2,\\
		Z&=& 1-e^2,\\
		V&=&p X+2 b (2-Z),\\
		W&=&b^2 Z^2+2 b p ((p+2) (1-Z)+p-2)+p^2 Y.
	\end{eqnarray}
	When $b=0$, we recover the results of Schwarzschild case.
	
	\subsection{Roots of the radial motion equation} \label{Zeros} 
	To obtain the roots of  $\mathcal{R}[r]$,  it can be reformulated as a polynomial in terms of $r$:
	\begin{eqnarray} \label{Rr0}
		- r^4 \mathcal{R}[r] /(1-\mathcal{E}^2) &=&  R[r]  \notag \\
		&=& r^4 +  A_3 r^3  +  A_2 r^2+  A_1 r^1+  A_0  \notag\\
		&=&(r-r_1)(r-r_2)(r-r_3)(r-r_4)=0 ,
	\end{eqnarray}
	where, 
	\begin{eqnarray}
		A_{0} &=&\frac{b \mathcal{J}  (\mathcal{J}+4 s_\parallel \mathcal{E})}{1 - \mathcal{E}^2} , \\
		A_{1} &=&-\frac{2 \mathcal{J} (\mathcal{J}+3 s_\parallel \mathcal{E})}{1-\mathcal{E}^2}, \\
		A_{2} &=&\frac{b +\mathcal{J}^2+2 \mathcal{J} s_\parallel \mathcal{E}}{1-\mathcal{E}^2},  \\
		A_{3} &=&-\frac{2 }{1-\mathcal{E}^2}.
	\end{eqnarray}
	We can obtain the remaining two roots $\{r_{3}, r_{4}\}$ as
	\begin{eqnarray}
		r_{3}&=& -\frac{1}{2} (A_3+r_1+r_2) \notag\\
		&&+\frac{\sqrt{r_1 r_2 (-4 A_0+r_1 r_2 (A_3+r_1+r_2)^2)}}{2 r_1 r_2},\label{r3}\\
		r_{4}&=&\frac{A_0}{ r_1 r_2 r_{3}}.\label{r4}
	\end{eqnarray}
	It can be verified that for bound orbits, we have
	\begin{eqnarray}
		r_1 r_2 (A_3+r_1+r_2)^2-4 A_0\approx \frac{4 p^4 (p-b) \left((1-b) p^3+b p^2 -b^2 p - b^3 e^2 \right)}{\left(1-e^2\right) \left(b^2 \left(e^2-1\right)+4 b p+(p-4) p^2\right)^2} + \mathcal{O}[s_\parallel] .
	\end{eqnarray}
	This value is usually greater than zero. Therefore, $\{r_{3}, r_{4}\}$ are all real roots.
	
	Expanding Eqs. \eqref{r3} and \eqref{r4} in terms of $s_\parallel$, we obtain the expressions of $\{r_{3}, r_{4}\}$, 
	\begin{widetext}
		\begin{eqnarray}
			r_{3}&=&\frac{p Z \left((p-b)^2+\sqrt{ U_1 }\right)}{p^2 V-W}+ s_\parallel \frac{ Z^2 \sqrt{W (p-b)}}{2 \left(p^2 V-W\right)^2}  \left(2 p \left(b (Z-2 p)+p^2\right)+\frac{ U_2 }{\sqrt{ U_1 }}\right)+ \mathcal{O}[s^2],\\
			r_{4}&=&\frac{b p (p-b)}{(p-b)^2+\sqrt{ U_1 }}  +  \frac{s_\parallel  b Z \sqrt{p-b}}{2 \left((p-b)^2+\sqrt{ U_1 }\right)^2 \left(p^2 V-W\right)}  \label{r3toep}\\
			&&  \bigg( 2 \sqrt{ U_3 } \left((p-b)^2+\sqrt{ U_1 }\right) (2 (p-2) p + b Z) \\
			&& -\sqrt{W} (p-b) \left(2 p \left(p^2 + b (Z-2 p)\right)+\frac{ U_2 }{\sqrt{ U_1 }}\right) \bigg) + \mathcal{O}[s^2],  \label{r4toep}
		\end{eqnarray}
	\end{widetext}
	where, 
	\begin{eqnarray}
		U_1&=&(p-b) \left(p^3-b (p-1) p^2-b^2 p-b^3 (1-Z)\right) ,\\
		U_2&=&2 p^5 - 2 b p^3 ((p-2) p+8-Z) +b^2 p^2 (16-p (Z-2)) \notag\\
		&&+2 b^3 p (p (Z-2)-3 Z)+b^4 Z^2 ,\\
		U_3&=& p^2 ((p-4) p+4 Z) -2 b p (p (Z-2)+2 Z) + b^2 Z^2.
	\end{eqnarray}
	From the above, it can be observed that when $b \leq 1$, both $U_1$ and $U_3$ are greater than zero outside the ISO, meanwhile,
	\begin{eqnarray}
		r_1>r_2>r_3>r_4 .
	\end{eqnarray}

	\section{Analytical solution}\label{Analytic}
	In this section, we use the Carter-Mino time $\lambda$ instead of the proper time $\tau$ for integration, with $d\tau/d\lambda=  r^2$ \cite{Global_Carter_1968,Perturbative_Mino_2003}.
	The main integrations of the equations of motion include 
	\begin{eqnarray}
		\lambda[r]-\lambda[r_2] &=&\int_{r_{2}}^{r}\frac{\mathrm{d}\bar{r} }{\sqrt{(1-\mathcal{E}^{2})R[\bar{r}]}}, \label{lambdar}\\
		t[r]-t[r_2] &=&\int_{r_{2}}^{r}\frac{\mathrm{d}\bar{r}}{\sqrt{(1-\mathcal{E}^{2})R[\bar{r}]}} \bar{r}^2 \left( \frac{\mathcal{E}}{f[\bar{r}]} +  \frac{\mathcal{J}f'[\bar{r}] s_\parallel}{2 \bar{r} f[\bar{r}]}\right),    \label{tr}\\
		\phi[r]-\phi[r_2] &=&\int_{r_{2}}^{r}\frac{\mathrm{d}\bar{r}}{\sqrt{(1-\mathcal{E}^{2})R[\bar{r}]}}  \bar{r}^2 \left( \frac{\mathcal{J}}{ \bar{r}^{2}} + \frac{  \mathcal{E} s_\parallel}{\bar{r}^{2}}  \right),    \label{phir}\\
		\psi[r]-\psi[r_2] &=&\int_{r_{2}}^{r}\frac{\mathrm{d}\bar{r}}{\sqrt{(1-\mathcal{E}^{2})R[\bar{r}]}}  \bar{r}^2 \left( \frac{\mathcal{E}\mathcal{J} }{\bar{r}^{2}+\mathcal{J}^{2}}  \right).  \label{psir}
	\end{eqnarray}
	
	\subsection{Integral}
	
	Through the integral in Eq.\eqref{lambdar}, the Carter-Mino time $\lambda$ as a function of the radial $r$ can be expressed as
	\begin{eqnarray}
		\lambda(r)-\lambda(r_2)&=&\frac{2 F_e[\chi,k^2]}{\sqrt{(1-\mathcal{E}^{2})(r_{1}-r_{3})(r_{2}-r_{4})}}, 
	\end{eqnarray}
	where 
	\begin{eqnarray}
		\sin\chi &=& \sqrt{\frac{(r_{1}-r_{3})(r-r_{2})}{(r_{1}-r_{2})(r-r_{3})}}, \\
		k^{2}&=&\frac{(r_{1}-r_{2})(r_{3}-r_4)}{(r_{1}-r_{3})(r_{2}-r_4)}.
	\end{eqnarray}
	
	Through the  indefinite integral form of Eqs. \eqref{tr}-\eqref{psir}, we can express $\{t, \phi, \psi\}$ as a function of the parameter $\chi[r]$ as 
	\begin{widetext}
		\begin{eqnarray}
			\tilde{T}_r[\chi]=&&\frac{1}{\sqrt{1-\mathcal{E}^{2}}\sqrt{(r_{1}-r_{3})(r_{2}-r_{4})}} \notag\\
			&&\bigg( \mathcal{E} (r_{1}-r_{3}) (r_{2}-r_{4})  E_e[\chi,k^{2}] +\frac{\mathcal{E} \sin[2 \chi] (r_{1}-r_{2})}{(-r_{1}-\cos[2 \chi] (r_{1}-r_{2})-r_{2}+2 r_{3})} \notag\\
			&&\left(\sqrt{(r_{1}-r_{3})(r_{2}-r_{4})-\sin[\chi]^{2}(r_{1}-r_{2})(r_{3}-r_{4})} \sqrt{(r_{1}-r_{3})(r_{2}-r_{4})}\right)\notag\\
			&&+\left(r_2-r_3\right)\left(r_1+r_2+r_3+r_{4}+4\right) \mathcal{E} \Pi_e[h_r,\chi,k^2]\\		
			&&+\frac{2  I_{nC}  (r_{2}-r_{3}) \Pi_e[h_{rn}, \chi ,k^{2}]}{(r_{2}-r_{n}) (r_{3}-r_{n})}+ \frac{2  I_{pC}  (r_{2}-r_{3}) \Pi_e[h_{rp}, \chi ,k^{2}]}{(r_{2}-r_{p}) (-r_{3}+r_{p})} +\frac{F_e[\chi ,k^{2}]}{(r_{3}-r_{n}) (r_{3}-r_{p})}\notag\\
			&&\Big(\mathcal{E}\left(8-2 b+r_{1} (-r_{2}+r_{3}) + r_{3} (4+r_{2}+r_{3})\right) (r_{3}-r_{n}) (r_{3}-r_{p})\notag\\
			&&-2 (r_{3}-r_{p})I_{nC}+2 (r_{3}-r_{n}) I_{pC}\Big) \bigg) ,\\
			\tilde{\Phi}_r[\chi]&=&
			\frac{2  F_e[\chi, k^{2}]  (\mathcal{J}+s_\parallel \mathcal{E})}{\sqrt{1-\mathcal{E}^2} \sqrt{\left(r_1-r_3\right) \left(r_2-r_4\right)}} ,\\
			\tilde{\Psi}_r[\chi]&=&\frac{ 2\mathcal{E}\mathcal{J} }{\sqrt{(r_{1}-r_{3})(r_{2}-r_{4})}\sqrt{1-{\mathcal E}^{2}}\left(r_{2}^{2}+{\mathcal{J}}^{2}\right)\left(r_{3}^{2}+{\mathcal{J}}^{2}\right)}\notag\\
			&&\Big( r_{3}{}^{2}\left(r_{2}{}^{2}+\mathcal{J}^{2}\right) F_e[\chi,k^{2}] \notag \\
			&&+(r_{2}-r_{3})\left(r_{2}r_{3}-\mathcal{J}^{2}\right) \mathcal{J} \mathrm{Im}\Big[\Pi_r \Big[\frac{(r_{1}-r_{2})(r_{3}-i\mathcal{J})}{(r_{1}-r_{3})(r_{2}-i\mathcal{J})},\chi,k^{2}\Big]\Big]  \notag \\
			&&+(r_{2}-r_{3})(r_{2}+r_{3})\mathcal{J}^{2}\mathrm{Re}\Big[ \Pi_r \Big[\frac{(r_{1}-r_{2})(r_{3}-i\mathcal{J})}{(r_{1}-r_{3})(r_{2}-i\mathcal{J})},\chi,k^{2}\Big]\Big]\Big) ,
		\end{eqnarray}
	\end{widetext}
	where,
	\begin{eqnarray}
		h_r&=& \frac{r_1-r_2}{r_1-r_3},\quad h_{rp}= h_r\frac{r_3-r_p}{r_2-r_p},\quad h_{rn}= h_r\frac{r_3-r_n}{r_2-r_n},\\
		I_{rC1}&=&8 \mathcal{E}-4 \mathcal{E} b +\mathcal{J} s_\parallel,  \quad I_{rC2}=-4 \mathcal{E} b  +  \mathcal{E} b^2-\mathcal{J} b s_\parallel, \\
		I_{pC} &=& \frac{1}{r_p-r_n} \left( I_{rC1} r_p + I_{rC2}\right), \quad I_{nC} = \frac{1}{r_p-r_n} \left( I_{rC1} r_n + I_{rC2}\right).\\
	\end{eqnarray}

	The above integration results are all within one period in radial direction. Following the spirit of \cite{Analytic_Meent_2020}, we can further obtain the following relationships:
	\begin{eqnarray}
		r[\lambda] &=& \frac{r_{3}(r_{1}-r_{2})\mathrm{sn}^{2}\left[K_e[k^2]q^{r}/\pi,k^2\right]-r_{2}(r_{1}-r_{3})}{(r_{1}-r_{2})\mathrm{sn}^{2}\left[K_e[k^2]q^{r}/\pi,k^2\right]-(r_{1}-r_{3})},\notag\\ 
		t[\lambda] &=&
		q^{t}+\tilde{T}_{r}\left[\mathrm{am}\left[\frac{q^{r}}{\pi}K_e[k^2],k^2\right]\right]-\frac{\tilde{T}_{r}[\pi]}{2\pi}q^{r},  \\
		\phi[\lambda]&=&
		q^{\varphi}+\tilde{\Phi}_{r}\left[\mathrm{am}\left[\frac{q^{r}}{\pi}K_e[k^2],k^2\right]\right]-\frac{\tilde{\Phi}_{r}[\pi]}{2\pi}q^{r}, \\
		\psi[\lambda]&=&
		q^{\psi}+\tilde{\Psi}_{r}\left[\mathrm{am}\left[\frac{q^{r}}{\pi}K_e[k^2],k^2\right]\right]-\frac{\tilde{\Psi}_{r}[\pi]}{2\pi}q^{r}, 
	\end{eqnarray}
	where,
	\begin{eqnarray}
		&&q^r=\Upsilon^r\lambda+q_0^r, \label{qr}\\
		&&q^{t}=\Upsilon^{t}\lambda+q_{0}^{t},\\
		&&q^{\phi}=\Upsilon^{\phi}\lambda+q_{0}^{\phi},\\
		&&q^{\psi}=\Upsilon^{\psi}\lambda+q_{0}^{\psi}. \label{qpsi}
	\end{eqnarray}
	
	The corresponding frequency is
	\begin{eqnarray}
		\Upsilon^{r} &=& \frac{\pi\sqrt{(1-\mathcal{E}^{2})(r_{1}-r_{3})(r_{2}-r_{4})}}{2K_e[k^2]}  ,\\
		\Upsilon^{t} &=&\frac{I_{pC}}{r_3-r_p} - \frac{I_{nC}}{r_3-r_n}+\frac{\mathcal{E}}{2}\left(-2 b+r_1 \left(r_3-r_2\right)+r_3 \left(r_2+r_3+4\right)+8\right) \notag\\
		&&+\frac{\left(r_1-r_3\right) \left(r_2-r_4\right) \mathcal{E} E_e[k^2]}{2 K_e[k^2]}+\frac{\left(r_2-r_3\right)}{2 K_e[k^2]} \bigg(\frac{2 I_{nC} \Pi_e[h_{rn},k^2]}{\left(r_2-r_n\right) \left(r_3-r_n\right)} \notag\\
		&& +\frac{2 I_{pC} \Pi_e[h_{rp},k^2]}{\left(r_2-r_p\right) \left(r_p-r_3\right)}+\left(r_1+r_2+r_3+r_4+4\right) \mathcal{E} \Pi_e[h_r,k^2]\bigg) ,\\
		\Upsilon^{\phi} &=& \mathcal{J} + \mathcal{E} s_\parallel ,\\
		\Upsilon^{\psi} &=& \frac{\mathcal{E} \mathcal{J} r_3^2 }{\mathcal{J}^2+r_3^2}+\frac{\mathcal{E} \mathcal{J}^2 \left(r_2-r_3\right) }{\left(\mathcal{J}^2+r_2^2\right) \left(\mathcal{J}^2+r_3^2\right) K_e[k^2]} \notag\\
		&&\bigg(\left(r_2 r_3-\mathcal{J}^2\right) \mathrm{Im} \left[\Pi _e\left[\frac{\left(r_1-r_2\right) \left(r_3-i \mathcal{J}\right)}{\left(r_1-r_3\right) \left(r_2-i \mathcal{J}\right)},k^2\right]\right]\notag\\
		&&+\mathcal{J} \left(r_2+r_3\right) \mathrm{Re} \left[\Pi _e\left[\frac{\left(r_1-r_2\right) \left(r_3-i \mathcal{J}\right)}{\left(r_1-r_3\right) \left(r_2-i \mathcal{J}\right)},k^2\right]\right]\bigg) .
	\end{eqnarray}
	It should be noted again that when $b=0$, the $s_\parallel$ part of $\Upsilon^{t}$ is not consistent with (52) in \cite{Analytic_Witzany_2024}, but it matches the result in their Supplemental Material. Our result is detailed in Appendix \ref{Doubt}.
	
	\begin{figure}[!htb]
		\centering
		\subfigure[{$s_\parallel=s_\perp=e=0.3$ and $p=10$}]{
			\label{orbitall1}
			\begin{minipage}[b]{0.35\textwidth}
				\includegraphics [width=1\textwidth]{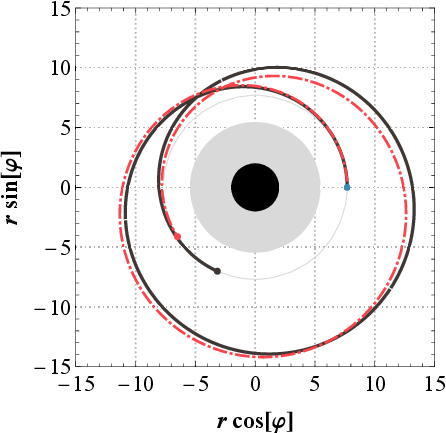}
				\includegraphics [width=1\textwidth]{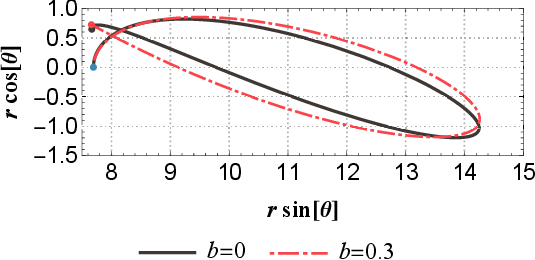}
		\end{minipage}}
		\subfigure[{$s_\perp=b=e=0.3$ and $p=10$}]{
			\label{orbitall2}
			\begin{minipage}[b]{0.35\textwidth}
				\includegraphics [width=1\textwidth]{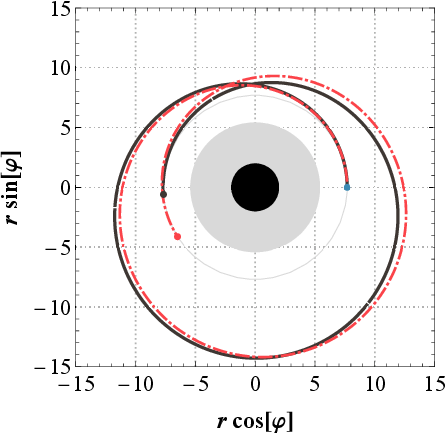}
				\includegraphics [width=1\textwidth]{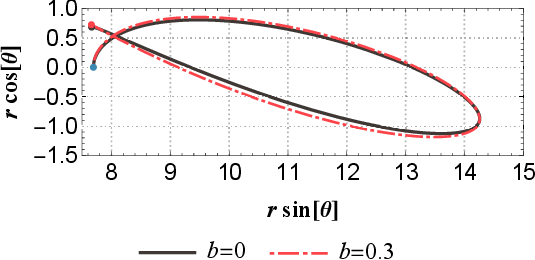}
		\end{minipage}}
		\caption{The effects of different value of tidal charge $b$ (left column) or  spin  $s_\parallel$ (right column) on the orbit in the angular-momentum-aligned coordinates. Corresponding to the black orbit, The central black circle and the gray one represent the black hole and the ISO range, respectively.}
		\label{orbitall}
	\end{figure}
	In Fig. \ref{orbitall}, we plot the orbit over one radial period using the analytical solutions \eqref{qr}-\eqref{qpsi}. The blue dot represents the starting point, while the red and black dots are the endpoints, but both at the periapsis. To display the effects of spin, we did not choose a small value of $s_\parallel$ and $s_\perp$. As seen in Fig. \ref{orbitall}, increasing in tidal charge $b$ reduces the periapsis precession angle, whereas increasing in spin enhances it.
	
	\section{Linear correction to the orbital frequencies}\label{FShift}
	The linear correction to the Mino frequency reads
	\begin{eqnarray}
		\Upsilon^{r,t,\phi}  &=& \Upsilon^{r,t,\phi}_{b} + s_\parallel \delta\Upsilon^{r,t,\phi}. 
	\end{eqnarray}
	Here, $\Upsilon^{r,t,\phi}_{b}$ represents the geodesic Mino frequency without considering the spin of the test particle. Additionally, we can use the earlier expressions to reorganize the energy $\mathcal{E}^2$ (Eq.\eqref{E2toep}), angular momentum $\mathcal{J}^2$ (Eq.\eqref{J2toep}), and the roots $r_3$ (Eq.\eqref{r3toep}) and $r_4$ (Eq.\eqref{r4toep}) as:
	\begin{eqnarray}
		\mathcal{E}^2  &=& \mathcal{E}_{b}^2 + s_\parallel \mathcal{E}_{s}^2 ,\\
		\mathcal{J}^2  &=& \mathcal{J}_{b}^2 + s_\parallel \mathcal{J}_{s}^2 ,\\
		r_3 &=& r_{b3} + s_\parallel r_{s3} ,\\
		r_4 &=& r_{b4} + s_\parallel r_{s4} .
	\end{eqnarray}
	The other expressions without spin can be written as $k_{b}^2$, $h_{br}$, $h_{brp}$, $h_{brn}$, etc. For example,
	\begin{eqnarray}
		k_{b}^2=\frac{(r_{b1}-r_{b2})(r_{b3}-r_{b4})}{(r_{b1}-r_{b3})(r_{b2}-r_{b4})}.
	\end{eqnarray}
	
	Therefore, $\Upsilon^{r}_{b}$ and its corresponding linear correction part $\delta\Upsilon^{r}$ can be described as
	\begin{eqnarray}
		\Upsilon^{r}_{b} &=& \frac{\pi  \sqrt{1-\mathcal{E}_b^2} \sqrt{\left(r_{b1}-r_{b3}\right) \left(r_{b2}-r_{b4}\right)}}{2 K_e[k_b^2]} ,\\
		\delta\Upsilon^{r} &=& -\frac{\pi  \sqrt{\left(r_{b1}-r_{b3}\right) \left(r_{b2}-r_{b4}\right)}}{4 \sqrt{1-\mathcal{E}_b^2} K_e[k_b^2]^2 \left(r_{b1}-r_{b4}\right) \left(r_{b3}-r_{b2}\right) \left(r_{b3}-r_{b4}\right)} \notag\\
		&&\Big(\left(\mathcal{E}_b^2-1\right) E_e[k_b^2] \left(r_{s4} \left(r_{b1}-r_{b3}\right) \left(r_{b3}-r_{b2}\right)+r_{s3} \left(r_{b1}-r_{b4}\right) \left(r_{b2}-r_{b4}\right)\right)    \notag\\
		&&-K_e[k_b^2] \left(r_{b1}-r_{b4}\right) \left(r_{b2}-r_{b3}\right) \left(\left(\mathcal{E}_b^2-1\right) \left(r_{s3}-r_{s4}\right)+\mathcal{E}_s^2 \left(r_{b3}-r_{b4}\right)\right) \Big) .
	\end{eqnarray}
	When $b=0$, our results goes back to the corresponding formula in Ref. \cite{Analytic_Witzany_2024} with $k_b^2$ term becomes $k_0^2$. Similarly, the Mino frequencies for $\phi$ and $\psi$ are
	\begin{eqnarray}
		\Upsilon^{\phi}_{b}   &=& \mathcal{J}_b ,\\
		\delta\Upsilon^{\phi} &=&\frac{\mathcal{J}_s^2}{2 \mathcal{J}_b}+\mathcal{E}_b ,\\
		\Upsilon^{\psi}_{b}   &=&\frac{\mathcal{E}_b \mathcal{J}_b r_{b3}^2 }{\mathcal{J}_b^2+r_{b3}^2}+\frac{\mathcal{E}_b \mathcal{J}_b^2 \left(r_{b2}-r_{b3}\right) }{\left(\mathcal{J}_b^2+r_{b2}^2\right) \left(\mathcal{J}_b^2+r_{b3}^2\right) K_e[k_b^2]} \notag\\
		&&\bigg(\left(r_{b2} r_{b3}-\mathcal{J}_b^2\right) \mathrm{Im} \left[\Pi _e\left[\frac{\left(r_{b1}-r_{b2}\right) \left(r_{b3}-i \mathcal{J}_b\right)}{\left(r_{b1}-r_{b3}\right) \left(r_{b2}-i \mathcal{J}_b\right)},k_b^2\right]\right]\notag\\
		&&+\mathcal{J}_b \left(r_{b2}+r_{b3}\right) \mathrm{Re} \left[\Pi _e\left[\frac{\left(r_{b1}-r_{b2}\right) \left(r_{b3}-i \mathcal{J}_b\right)}{\left(r_{b1}-r_{b3}\right) \left(r_{b2}-i \mathcal{J}_b\right)},k_b^2\right]\right]\bigg) .
	\end{eqnarray}
	Then, we can obtain the  the nodal precession $\nu_{nodal}$ and the periastron precession  $\nu_{peri}$ as 
	\begin{eqnarray}
		\nu_{nodal} &=& 2 \pi \left(\frac{\Upsilon^{\phi}_{b}}{\Upsilon^{\psi}_{b}}-1\right), \label{nunodal}\\
		\nu_{peri} &=& \nu^b_{peri} + \nu^s_{peri} s_\parallel  = 2 \pi  \left(\frac{\Upsilon^{\phi}_{b}}{\Upsilon^{r}_{b}}-1\right) +  2 \pi  \frac{ \Upsilon^{r}_{b} \delta\Upsilon^{\phi} -\delta\Upsilon^{r}  \Upsilon^{\phi}_{b} }{\left(\Upsilon^{r}_{b}\right)^2} s_\parallel.  \label{nuperi}
	\end{eqnarray}
	From Eq.\eqref{nunodal}, we consider the spinless part of  $\nu_{nodal}$. This is because, from Eq.\eqref{psitau}, nodal precession is only an $\mathcal{O}[s]$ order effect. In contrast, periastron precession requires up to the linear order in spin.
	
	\begin{figure*}[!htb]
		\centering
		\subfigure[{$e=0.3$ and $p=10$}]{
			\includegraphics[width=0.4\textwidth]{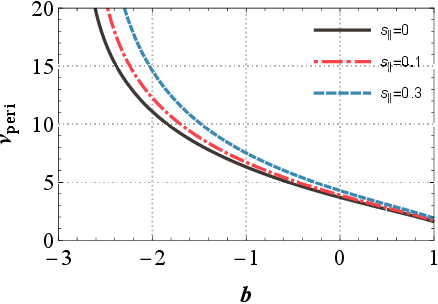}}
		\subfigure[{$e=0.3$ and $p=10$}]{
			\includegraphics[width=0.4\textwidth]{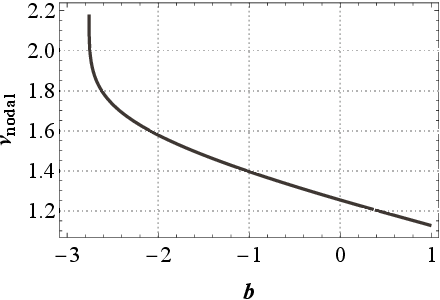}}
		\caption{Nodal and periastron precession under different tidal charge $b$ and spin $s_\parallel$.}
		\label{FShiftFig}
	\end{figure*}
	The nodal and the periastron precession for different values of $b$ and $s_\parallel$ is illustrated in Fig. \ref{FShiftFig}.
	From Fig. \ref{FShiftFig}, it can be seen that periapsis precession increases as the tidal charge $b$  decreases and $s_\parallel$ increases. This is consistent with the conclusions from Fig. \ref{orbitall}.

	\section{Conclusion}\label{conclusion}
	In this paper, we considered the  explicit orbital solutions for the motion of a spinning test particle in the brane-world spacetime. Along the line of \cite{Analytic_Witzany_2024}, we first analyzed the equations of motion for the test spinning particle, then provided the evolution of the spin components through parallel transport. Next, we derived the Keplerian parameter expressions for the energy and angular momentum, as well as the four roots of the radial velocity. As a result, we presented the elliptical integral calculation of the equations of motion in the Carter-Mino time and obtained the linear shift of the orbital frequency. The influences of different value of tidal charge $b$ or  spin  $s_\parallel$ on the orbit in the angular-momentum-aligned coordinates has depicted in Fig. \ref{orbitall}. Moreover, we found that both the nodal precession and the periastron precession become larger when the tidal charge $b$ of brane-world spacetime becomes smaller. For the periastron precession, the effect will be further amplified as the spin increases.
	
	There are a lot of works that deserve further investigation. For example, it is interesting to extend these results to the case of Reissner-Nordstrom spacetime with neutral or charged test particles. In such a scenario, the test particle will not only be subject to the spin-curvature force but also the electromagnetic interaction force. Additionally, we anticipate using the conclusions derived from this method to study gravitational waveforms in EMRI systems.
	
	\begin{acknowledgments}
		Some of this work make use of xAct\cite{xAct_Martin_2017}.
		This work is supported by National Natural Science Foundation of China (NSFC) with Grants No. 12275087. 
	\end{acknowledgments}
	
	\appendix

	\section{Effective potential and bound orbits}
	The effective potential $V_{eff}$ can be defined as \cite{Relativity_Rindler_2006,Periodic_Deng_2020}
	\begin{eqnarray}
		V_{eff} &=& \mathcal{E}^2 - \dot {r}^{2}  \notag\\
		&=&  \frac{(\mathcal{J}^{2} + r^2) f[r]}{r^{2}} +  \frac{\mathcal{J} \mathcal{E} ( 2f[r] - rf^{\prime}[r] )  s_{\parallel} }{r^2}.
	\end{eqnarray}
	We can see that, unlike the spinless case, the inclusion of spin effects introduces a coupling term $\mathcal{J} \mathcal{E} s_{\parallel}$ and implies that changing the energy $\mathcal{E}$ does not simply translate the effective potential up or down. This feature makes finding analytical solutions for marginally bound orbits (MBO) and innermost stable circular orbits (ISCO) more complex.
	
	For MBO, the condition is 
	\begin{eqnarray}
		V_{eff}[r_{MBO}] = 1, \quad
		V'_{eff}[r_{MBO}] = 0, 
	\end{eqnarray}
	while for ISCO, the condition becomes
	\begin{eqnarray}
		V_{eff}[r_{ISCO}] =  \mathcal{E}^2, \quad
		V'_{eff}[r_{ISCO}] = 0, \quad
		V''_{eff}[r_{ISCO}] = 0.
	\end{eqnarray}
	
\begin{figure*}[!htb]
		\centering
		\subfigure[{$s_\parallel=0.2$, $b=-1$}]{
			\includegraphics[width=0.31\textwidth]{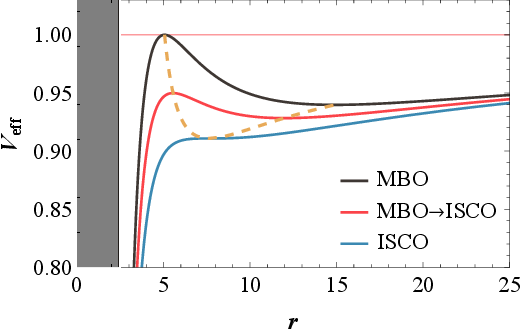}}
		\subfigure[{$s_\parallel=0.2$, $b=0$}]{
			\includegraphics[width=0.31\textwidth]{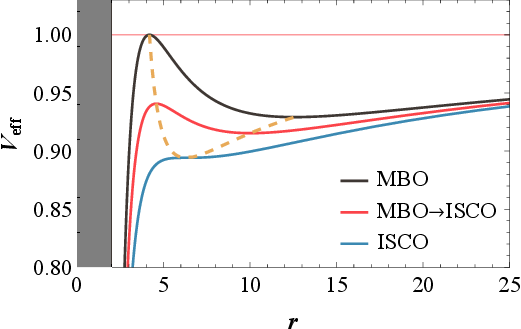}}
		\subfigure[{$s_\parallel=0.2$, $b=1$}]{
			\includegraphics[width=0.31\textwidth]{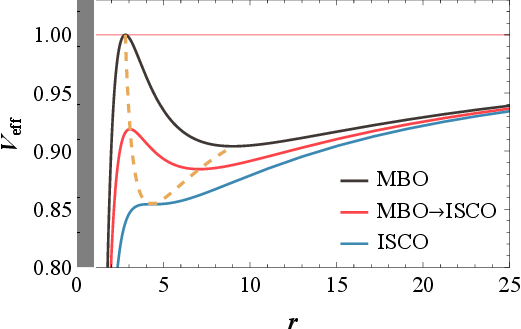}}
		\subfigure[{$s_\parallel=-0.2$, $b=-1$}]{
			\includegraphics[width=0.31\textwidth]{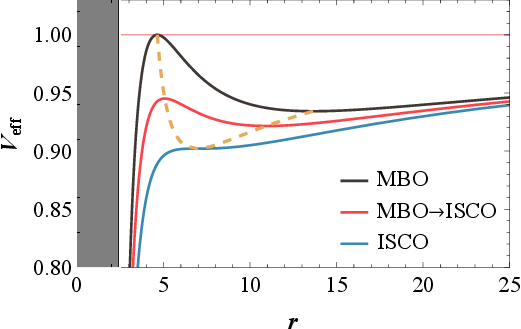}}
		\subfigure[{$s_\parallel=-0.2$, $b=0$}]{
			\includegraphics[width=0.31\textwidth]{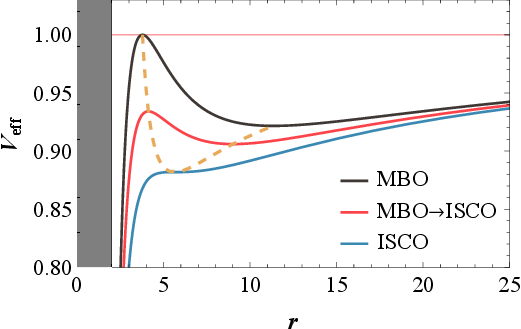}}
		\subfigure[{$s_\parallel=-0.2$, $b=1$}]{
			\includegraphics[width=0.31\textwidth]{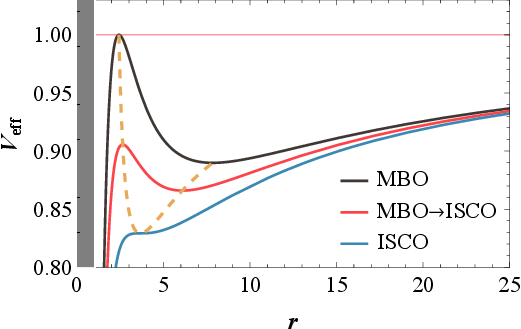}}
		\caption{The effective potential $V_{eff}[r]$  for different values of $s_\parallel$ and $b$. The Energy $\mathcal{E}$ and  angular momentum $\mathcal{J}$ varies from $\{\mathcal{E}_{ISO} , \mathcal{J}_{ISO}\}$ to $\{\mathcal{E}_{MBO} , \mathcal{J}_{MBO}\}$ from bottom to top.
			The dashed orange line is located at the extremal points of $V_{eff}[r]$. The gray area on the left represents the part that inside the event horizon.}
		\label{Veff}
	\end{figure*}

	The numerical results are shown in Fig. \ref{Veff}.
	From Fig. \ref{Veff}, the radius of ISO (MBO) decreases as the tidal charge $b$ increases while as the $s_\parallel$ increases, the radius of  ISO (MBO)  increases.
	
	On the other hand, we have the expressions for the roots in Sec. \ref{Zeros}. Therefore, ISO can also be found by setting $r_2 = r_3$. However, even in the absence of spin effects, the analytical result for the ISO in the brane-world remains quite complex.
	\begin{figure}[!htb]
		\includegraphics [width=0.45\textwidth]{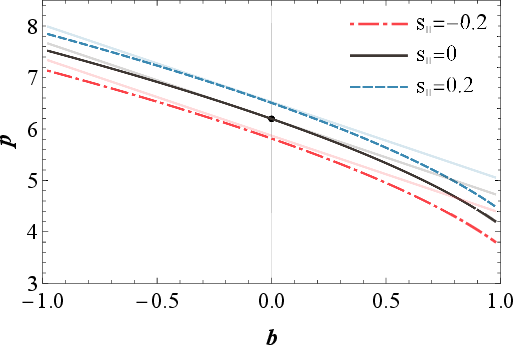}
		\caption{The semilatus rectum $p$ of the ISO as functions of
			the tidal charge $b$ for different values of $s_\parallel$,  with $e = 0.1$.}
		\label{pISO}
	\end{figure}
	Hence, in Fig. \ref{pISO}, we obtained  the semilatus rectum $p$ of the ISO numerically. Then $\{b, s_\parallel, e\}$ as small quantities, we can obtain the  approximation
	\begin{eqnarray}
		p = 6+2 e-\frac{3 b}{2}+2 \sqrt{\frac{2}{3}} s_\parallel .
	\end{eqnarray}
	As Fig. \ref{pISO} shows, when  $e$, $b$ and $s_\parallel$ approach to zero, the light line representing the approximate result approaches the numerical one.
	This approximate result is consistent with Eq.(B19) in \cite{Conservative_Favata_2011}. It can be seen that the $p$ of the ISO decreases with increasing  $b$ and decreasing $s_\parallel$. This is in contrast to the conclusion in \cite{Innermost_Zhang_2018}, where the ISCO ($e=0$) decreases with increasing spin $s$. This discrepancy arises because the second leg in the Marck tetrad has an opposite sign compared to the second basis in the tetrad used in \cite{Innermost_Zhang_2018}.
	
	\section{Formulae of integrals of radial motion}\label{Elliptic}
	$F_e[\chi, k^2]$, $E_e[\chi, k^2]$, and $\Pi_e[n, \chi, k^2]$ are the elliptic integrals of the first, second, and third kinds, respectively, as follows	\cite{Innermost_Zhang_2018}. 
	\begin{eqnarray}
		F_e[\chi,k^2] &=& \int_0^\chi (1-k^2 \sin^2[\theta])^{-1/2} d \theta  ,\\
		E_e[\chi,k^2] &=& \int_0^\chi (1-k^2 \sin^2[\theta])^{1/2} d \theta  ,\\
		\Pi_e[n, \chi, k^2] &=& \int_0^\chi (1-k^2 \sin^2[\theta])^{-1/2} (1-n \sin^2[\theta])^{-1} d \theta .
	\end{eqnarray}
	The corresponding complete elliptic integrals are given by
	\begin{eqnarray}
		K_e[k^2] &=& F_e[\pi/2 ,k^2], \\
		E_e[k^2] &=& E_e[\pi/2 ,k^2], \\
		\Pi_e[n, k^2] &=& \Pi_e[n, \pi/2 , k^2].
	\end{eqnarray}
	The inverses of the $F_e[\chi,k^2]$ is 
	\begin{eqnarray}
		F_e[\mathrm{am}[y,k^2],k^2]  &=& y ,\\
		\mathrm{sn}[y,k^2] &=& \sin[\mathrm{am}[y,k^2]].
	\end{eqnarray}
	
	\section{The result for Schwarzschild case}\label{Doubt}
	In the Schwarzschild case, we obtained the following result as
	\begin{eqnarray}
		\Upsilon^{t}_{Sh} &=&\frac{\mathcal{E}}{2 K_e[k^2]}\bigg(\frac{2 \left(r_1-r_3\right) \left(r_2-r_3\right)+r_3 \left(r_1 \left(r_3-r_2\right)+r_3 \left(r_2+r_3\right)\right)}{r_3-2} K_e[k^2] \notag\\
		&&+ \left(r_2-r_3\right) \left(r_1+r_2+r_3+4\right) \Pi _e\left[\frac{r_1-r_2}{r_1-r_3},k^2\right] +r_2 \left(r_1-r_3\right) E_e[k^2]\notag\\
		&& + \frac{16 \left(r_2-r_3\right)}{\left(r_2-2\right) \left(2-r_3\right)}\Pi_e\left[\frac{\left(r_1-r_2\right) \left(r_3-2\right)}{\left(r_1-r_3\right) \left(r_2-2\right)},k^2\right]\notag\\
		&&-\frac{\mathcal{J} s_\parallel}{\mathcal{E}}
		\left(\frac{2 \left(r_3-r_2\right)}{\left(r_2-2\right) \left(2-r_3\right)} \Pi_e\left[\frac{\left(r_1-r_2\right) \left(r_3-2\right)}{\left(r_1-r_3\right) \left(r_2-2\right)},k^2\right]+\frac{2 K_e[k^2]}{ \left(2-r_3\right)}\right)
		\bigg).
	\end{eqnarray}

	\bibliographystyle{unsrt}
	
\end{CJK}
\end{document}